\documentclass{aa}
\usepackage[varg]{txfonts}
\usepackage[utf8]{inputenc}
\usepackage{comment}
\usepackage{amsmath}
\usepackage{graphicx}
\usepackage{gensymb}
\usepackage[colorlinks=true,linkcolor=blue,citecolor=blue]{hyperref}
\usepackage{xspace}
\usepackage{chronology}
\usepackage{geometry}  
\usepackage{lipsum}    

\newcommand{\srii}{\ion{Sr}{ii}\xspace}
\newcommand{\sriii}{\ion{Sr}{iii}\xspace}
\newcommand{\yii}{\ion{Y}{ii}\xspace}
\newcommand{\zrii}{\ion{Zr}{ii}\xspace}

\newcommand{\laiii}{\ion{La}{iii}\xspace}
\newcommand{\ceiii}{\ion{Ce}{iii}\xspace}
\newcommand{\rprocess}{\textit{r}-process\xspace}

\usepackage{natbib}
\bibpunct{(}{)}{;}{a}{}{,} 

\begin{document}

\title{Kilonova evolution --- the rapid emergence of spectral features} 
\authorrunning{Sneppen et al.}

\author{Albert Sneppen\inst{\ref{addr:DAWN},\ref{addr:jagtvej}},
Darach Watson\inst{\ref{addr:DAWN},\ref{addr:jagtvej}},
James H. Gillanders\inst{\ref{addr:oxford}} \and
Kasper E. Heintz\inst{\ref{addr:DAWN},\ref{addr:jagtvej}} 
}

\institute{Cosmic Dawn Center (DAWN)\label{addr:DAWN}
\and
Niels Bohr Institute, University of Copenhagen, Jagtvej 128, DK-2200, Copenhagen N, Denmark \label{addr:jagtvej} 
\and
Astrophysics sub-Department, Department of Physics, University of Oxford, Keble Road, Oxford, OX1 3RH, UK\label{addr:oxford}
}

\date{Received date /
Accepted date }

\abstract{
      Kilonovae (KNe) are one of the fastest types of optical transients known, cooling rapidly in the first few days following their neutron-star merger origin. 
      We show here that KN spectral features go through rapid recombination transitions, with features due to elements in the new ionisation state emerging quickly. Due to time-delay effects of the rapidly-expanding KN, a `wave' of these new features passing though the ejecta should be a detectable phenomenon. In particular, isolated line features will emerge as blueshifted absorption features first, gradually evolving into more pronounced absorption/emission P\,Cygni features and then pure emission features. 
      In this analysis, we present the evolution of the individual exposures of the KN AT2017gfo observed with VLT/X-shooter that together comprise X-shooter's first epoch spectrum (1.43\,days post-merger). We show that the spectra of these `sub-epochs' show a significant evolution across the roughly one hour of observations, including a decrease of the blackbody temperature and photospheric velocity. The early cooling is even more rapid than that inferred from later photospheric epochs, and suggest a fixed power-law relation between temperature and time cannot capture the data. 
      The cooling constrains the recombination-wave, where a \srii interpretation of the AT2017gfo $\sim1\,\mu$m feature predicts both a specific timing for the feature emergence and its early spectral shape, including the very weak emission component observed at about 1.43\,days. This empirically indicates a strong correspondence between the radiation temperature of the blackbody and the ejecta's electron temperature. Furthermore, this reverberation analysis suggests that temporal modelling is important for interpreting individual spectra and that higher cadence spectral series, especially when concentrated at specific times, can provide strong constraints on KN line identifications and the ejecta physics. Given the use of such short-timescale information, we lay out improved observing strategies for future KN monitoring. \newline
        }
\keywords{}

\maketitle
\section{Introduction}
The temporal data on the kilonova AT2017gfo has provided several key insights into its evolution and composition. First, the bolometric luminosity follows a powerlaw-like decay, expected for a large ensemble of rapid neutron-capture process (\rprocess) isotopes \citep[e.g.][]{Metzger2010,Wu2019}. 
Second, follow-up analyses of the spectral energy distribution (SED) have identified P~Cygni features proposed to be formed by \rprocess elements, the first and most obvious being \srii \citep{Watson2019}, but also potentially \laiii and \ceiii \citep{Domoto2022}, as well as \yii \citep{Sneppen2023_yttrium}. These P~Cygni profiles change across epochs because they trace the receding photospheric velocity. They also potentially probe differing ionisation states and abundances through the ejecta. This means that more complete radiative transfer modelling of the spectrum can constrain changes in abundances and velocities as a function of time \citep{Gillanders2022,Vieira2023arXiv,Vieira2023}. Lastly, the continuum flux (across early epochs) is remarkably similar to a blackbody in terms of both normalisation and spectral shape \citep[e.g.][]{Sneppen2023_bb} with especially the NIR continuum being challenging to reproduce for current numerical simulations of mergers \citep[e.g.][]{Domoto2022,Collins2023b}. 

The nightly Very Large Telescope (VLT) spectra taken with the X-shooter spectrograph of the kilonova AT2017gfo associated with the gravitational wave event, GW170817, provide the highest-quality information yet available for studying the rapidly evolving electromagnetic output from the merger of neutron-stars \citep{Pian2017,Smartt2017}.  Due to observational constraints, spectra were mostly available with a cadence around 24 hours, limiting the temporal analysis to a time-step which is comparable to the time since merger for the early epochs. Temporally important earlier spectra were obtained at telescopes in Chile, Australia and South Africa \citep{Andreoni2017,McCully2017,Shappee2017,Buckley2018}, though not with the UV-NIR spectral coverage of the X-shooter spectra. While we return to these earlier spectra in a companion paper (Sneppen et~al. in~preparation), each of these `daily' X-shooter spectra is in fact a composite of several individual exposures taken over approximately an hour. For standard reductions these exposures are generally analysed together, but this is not required, given the brightness of the early kilonova, and the mildly relativistic velocities which smear the spectral line data over tens of thousands of km/s and obviate the need for high S/N at high spectral resolution. Conversely, given the rapid kilonova evolution it is informative to examine these exposures individually. Therefore, in Sect.~\ref{sec:reduction}, we re-reduce each exposure of the VLT, epoch~1 X-shooter spectrum of AT2017gfo (taken on 18--19 August 2017, 1.4 days post-merger). The reduced sub-epoch spectra are made publicly available and are intended as an extension and addition to the standard reduction presented in \citet{Pian2017,Smartt2017}. In Sect.~\ref{sec:delta}, we examine the physical information concealed in the differences between these sub-epochs. In Sect.~\ref{sec:emergence}, we show that on timescales of order several hours, reverberation effects dictate that KN spectral features will emerge rapidly and as blueshifted absorption components first. Lastly, in Sect.~\ref{sec:cadence} given the strong physical constraint residing in the temporal evolution, we discuss the optimal observing pattern, cadence and timing for future KNe. 

\begin{figure*}
    \centering
    \includegraphics[width=\linewidth,viewport=12 20 825 520, clip=]{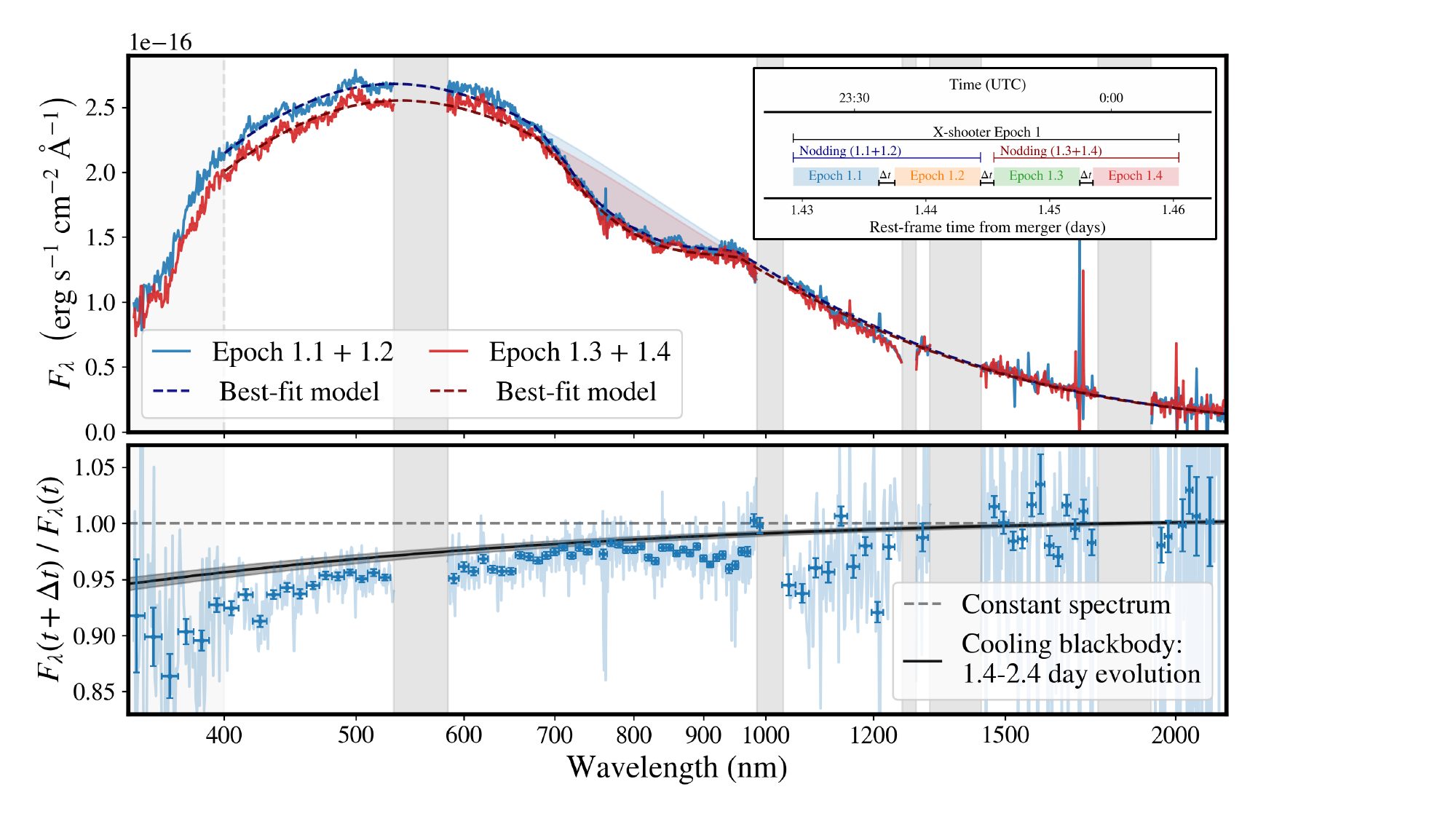}
    \caption{Top panel: X-shooter spectra of the first (blue) and second (red) pair of exposures from epoch~1. The timeline of exposures is shown in the the top-right inset). Corresponding best-fit models are shown with dashed lines. The spectral shapes are similar, but the later exposures are less luminous with a lower blackbody temperature. Shaded bars indicate regions with strong tellurics and poorly-constrained or noisy regions at the edge of the UVB, VIS and NIR arms of the spectrograph. Lower panel: The flux ratio as a function of wavelength (with the binned weighted mean overlaid), which shows a large fractional change in flux for the UV-arm, a gradual evolution across the VIS-arm and minor and somewhat noisy evolution in the NIR-arm. The cooling blackbody prediction is taken from the epoch 1--2 evolution from \citet{Sneppen2023}. The data match the model reasonably well overall. However, the data seem to require a stronger temperature decline at this early time than the model prediction based on the overall epoch~1--2 evolution.} 
    \label{fig:Epoch 1}
\end{figure*}

\section{Sub-epoch reduction and predicted variation}\label{sec:reduction}
In the following, we outline the data reduction and observational difference between the individual exposures of AT2017gfo made during epoch~1 (1.43\,days post-merger). For the first observing night, the medium-resolution, ultraviolet (320\,nm) to near-infrared (2480\,nm) spectrograph X-shooter took four exposures in nodding mode \citep{Pian2017}, which we label epochs 1.1, 1.2, 1.3 and 1.4. With 10 minute exposure-times and additional telescope overhead, these exposures monitor the evolution of the transient over about one hour. We focus on this early epoch, because it has the highest ratio of exposure timescale to the total time passed since the merger. The four exposures are reduced in nodding mode for the first two and latter two exposures ($t_{\rm diff} \approx 25$\,minutes, see Fig.~\ref{fig:Epoch 1}, top panel). We also reduced all four exposures individually in stare-mode, which shows the same temporal trend as the nodding-mode reduction. However, as the NIR background subtraction is difficult in stare-mode, we here focus on the nodding mode constraints. 




The spectral reduction pipeline follows those previously detailed in earlier publications \citep{Pian2017,Smartt2017}, where the X-Shooter pipeline supplied by ESO is used for flat-fielding, order tracing, rectification, and wavelength calibration. The flux calibration of the X-shooter spectra is broadly consistent with contemporaneous photometry from GROND and EFOSC2 \citep{Smartt2017}; DECAM \citep{Cowperthwaite2017}; LDSS \citep{Shappee2017}; Swope \citep{Coulter2017} and VIRCAM  \citep{Tanvir2017}. Therefore we follow the convention in \citet{Pian2017}, \citet{Watson2019} and \citet{Sneppen2023_H0} and do not artificially scale the spectra to any specific photometric data as these have significant internal scatter. We note, that the ``contemporaneous'' photometric fluxes commonly used for calibration are actually taken over a timescale of two hours, which should be considered a sizeable systematic effect in the reductions, given the rapid nature of the kilonova evolution at this time.

\subsection{Atmospheric observing conditions}\label{sec:atmospheric}

On the timescale of the sub-epochs, the atmospheric observing conditions can vary and affect the reduction output in three notable ways. 

First, the transmission curves and, by extension, the required telluric corrections, may change. Wavelengths with low atmospheric transmission, such as large parts of the NIR, are most susceptible to these changes, and a minor increase in transmission, for example, could lead to prominent spikes of the telluric-corrected flux. These spikes will be narrow (like the typical width of atmospheric absorption lines) and statistically weak, as the actual observed flux-count is small. To determine the impact of this effect, we find both the typical telluric correction estimated from the standard star (taken at comparable airmass immediately following the last exposure) and derive telluric corrections for each sub-epoch individually using \texttt{Molecfit} \citep{Smette2015,Kausch2015}. \texttt{Molecfit} was run using the standard, well-tested configuration and wavelength-ranges (1.12-1.13, 1.47-1.48, 1.80-1.81, 2.06-2.07$\mu m$) for fitting described in \cite{Kausch2015} -- albeit not including the wavelength-band around 2.35$\mu$m in the fit due to the weak signal at these wavelengths. These wavelengths-intervals are standard as they in combination probe the various atmospheric molecules. Because the standard star was only taken at the end of the observing sequence, only \texttt{Molecfit} can constrain the changes in telluric corrections over the individual exposures. Ultimately the difference between the two methods proves minor for the continuum, although fewer artificial and narrow peaks are seen in the intermittent telluric forests redward of $1300$\,nm when using \texttt{Molecfit}. 

Second, atmospheric dispersion for objects at low inclination is sizeable and chromatic. However, given the atmospheric dispersion correctors on X-shooter and the parallactic angle of the slit, we can verify that the spectral trace shows no skew with wavelength in any of the exposures.

Lastly, the seeing conditions were relatively favourable across the first epoch. The slit-width was $1''$, $0.9''$ and $0.9''$ for UVB, VIS and NIR, the airmass was in the range from 1.42-1.76, which corresponds to a seeing of $0.58''-0.74''$ ($0.59''-0.70''$) at 500\,nm for AT2017gfo (standard star). If the PSF width had become comparable to the slitwidth, slitloss becomes significant and must be accounted for, otherwise it would lead to an underestimation of the normalisation. Due to the wavelength-dependence of seeing, there would also be an increasing slitloss towards shorter wavelengths (which would lead to underestimation of the temperature). To establish an accurate flux calibration, slitloss corrections were calculated using the theoretical wavelength-dependence on seeing with the average seeing FWHM in each exposure \citep{Fried1966}, identical to the previous reductions \citep[e.g.][]{Pian2017} and further detailed in \cite{Selsing2019}. The slit losses are specifically obtained by integrating this synthetic point-spread-function over the width of the slits. However, the seeing in epoch~1 is of high enough quality that not accounting for slitloss corrections provides a similar wavelength-dependent spectral evolution between sub-epochs. Indeed, for all wavelengths $\lambda > 400$\,nm the resulting spectral ratio (see Fig. \ref{fig:Epoch 1}) is indistinguishable with and without slit-loss corrections, indicating the robustness of the reduction to this systematic effect. In the UV below 400\,nm a subtle difference of a few percent is applicable, but this small effect is inconsequential as it is minor and remains below the spectral range of focus in this analysis.




\subsection{Predicted kilonova cooling}\label{sec:predicted}

In a KN atmosphere, the temporal evolution of temperature is set by the competition between adiabatic cooling, radiative cooling and heating from the decay of \rprocess elements. The decays from an ensemble of isotopes is predicted to provide a heating rate well-approximated by a single power-law \citep{Metzger2010}, which is observationally supported by the AT2017gfo lightcurve, where the bolometric luminosity from 1 to 6 days post-merger follows a temporal (\(t\)) powerlaw-like decay, $L_{\rm bol} \propto t^{-0.95 \pm 0.06}$ \citep{Waxman2018}. The spectra of AT2017gfo in early epochs is observed to be very similar to a blackbody as quantified in \cite{Sneppen2023_bb}. In this case, for a spherical photosphere the wavelength-specific luminosity is \(L_{\lambda}^{\rm BB}(t) = 4 \pi R_{\rm ph}(t)^2 \pi B(\lambda,T(t))\), where \(R_{\rm ph}(t)\) is the photospheric radius, \(B(\lambda,T)\) is the Planck function with temperature, \(T\), at wavelength \(\lambda\). The ratio of fluxes at different times for a simple blackbody cooling model is thus merely the ratio of 1) Planck functions of different temperatures and 2) emitting areas. We deliberate on each of these in the following.

The spectral energy distribution ratio is quite sensitive to the evolution in temperature since we sample the Rayleigh-Jeans tail (i.e.\ $B_{\lambda \gg h c /(k_bT)}  \propto T$) which has a linear scaling of flux with temperature, the peak (where the flux-ratio for blackbodies is proportional to $T^{5}$), and the exponentially growing discrepancy between the Wien tails. A commonly used prescription for the temperature, $T$, as a function of the time $t$ since the merger is a powerlaw relation \citep[e.g.][]{Drout2017,Waxman2018}:
\begin{equation}
    T(t) = T_{\rm ref} \left(\frac{t}{t_{\rm ref}}\right)^{-\alpha}
    \label{eq:1}
\end{equation}
Here, $\alpha$ is the cooling powerlaw index and $T_{\rm ref}$ is the reference temperature at some reference time, $t_{\rm ref}$. Fitting blackbody models to photometry and spectra from 1 to 6 days post-merger suggests the observed temperature has a power-law decline with $\alpha=0.54 \pm 0.02$ \citep[e.g.][]{Waxman2018}. However, the temperature evolution may be more complex than a single powerlaw prescription. Indeed, more advanced spectral modelling which, along with a blackbody, also takes account of the $1\,\mu$m P~Cygni and the observed NIR emission lines, found a single powerlaw does not accurately describe the temperature evolution \citep{Sneppen2023}. It is cooling more rapid initially ($\alpha\approx0.62$) at epochs 1--2 (1.4--2.4 days post-merger), and slower ($\alpha\approx0.4$) at epochs 2--3 (2.4--3.4 days post-merger). 

The emitting area evolves subtly between the exposures. The atmosphere expands radially with $R(t) \propto t$ for homologous expansion. However, as the outer surface becomes increasingly optically thin, the photospheric boundary will recede deeper into the ejecta, so $R_{\rm ph}(t) \propto t^{ \beta}$ with $\beta<1$. Fitting blackbody models from 1--6 days post-merger suggests $\beta=0.61\pm0.05$ \citep{Waxman2018}. However, the \(1\,\mu\)m P~Cygni velocity and the Doppler-corrected blackbody-velocity from epochs 1--2 indicate a smaller recession effect with $\beta\approx0.75$ \citep{Sneppen2023}. This estimate improves on the previous blackbody velocity by accounting for the relativistic corrections \citep[see][]{Sneppen2023_bb} and by modelling the observed features in the spectrum (including \(1\,\mu\)m P~Cygni and NIR features, see \citealt{Watson2019}). Changing the emitting area predominantly produces an achromatic shift in the normalisation of the spectrum. Additionally, higher-order chromatic (albeit for this analysis negligible) effects may follow from a change in the velocity, such as different light-travel time-delays and relativistic corrections \citep[as quantified in][]{Sneppen2023_bb}. 

Ultimately, the evolution in epochs~1--2 provides the temporally nearest constraints on the predicted spectral evolution over the first epoch exposures. We will therefore define this model as the cooling blackbody prediction, though, a priori, one might expect a slightly faster evolution during epoch~1 itself than the transition from epoch~1--2 would give.

\begin{figure*}
    \centering
    \includegraphics[width=\linewidth]{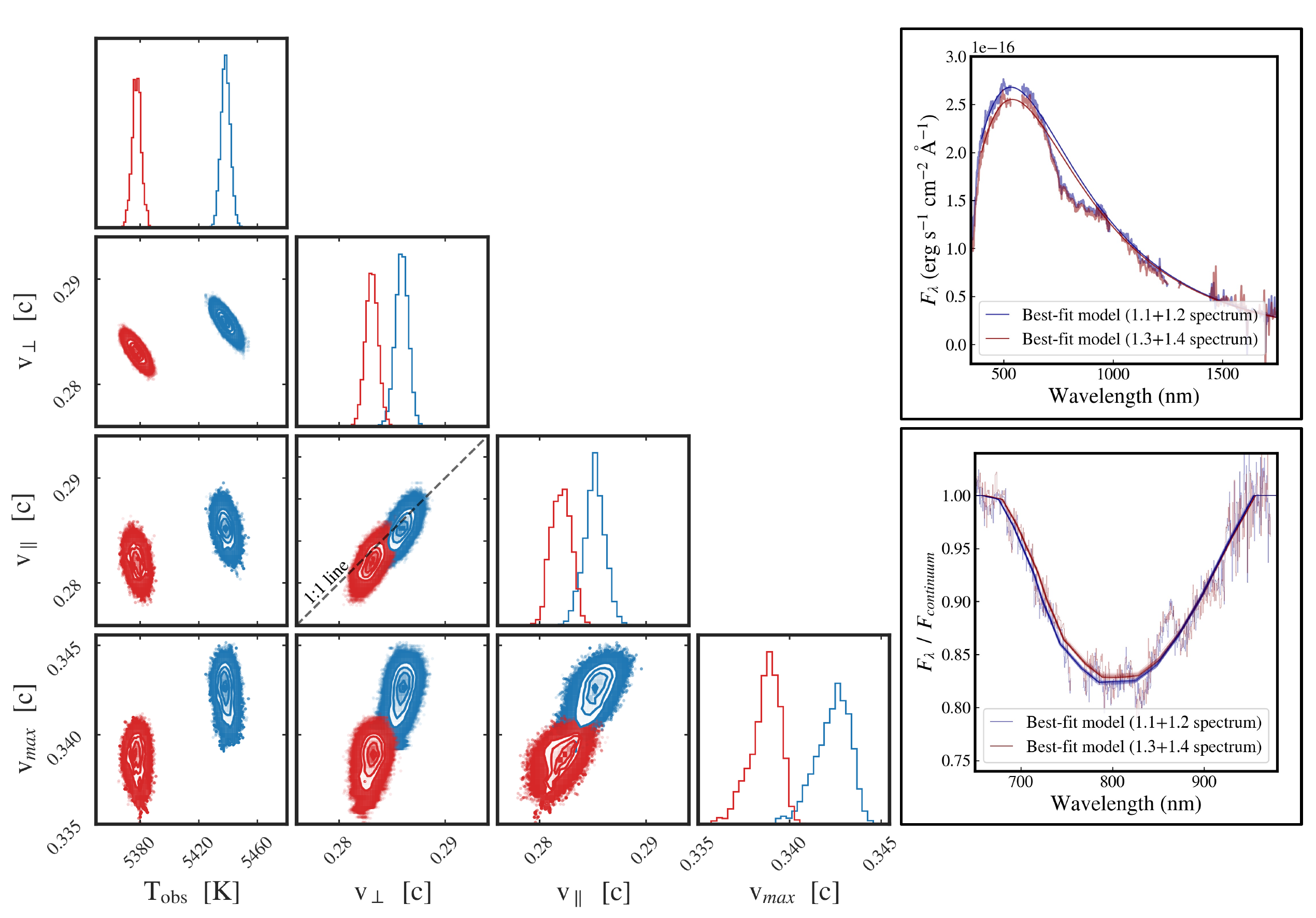}

    
    \caption{Corner-plot indicating the posterior probability distributions of key parameters from fitting the spectra in epoch 1.1+1.2 (blue) and epoch 1.3+1.4 (red). The fit parameters are (1) the best-fit observed blackbody temperature, $T_{\rm obs}$, (2) the cross-sectional velocity, $v_\perp$, derived from the blackbody normalisation, (3) the photospheric velocity of the $1\,\mu$m P~Cygni feature, $v_\parallel$, and (4) the outer ejecta velocity inferred from the $\,1\mu$m P~Cygni feature, $v_{\rm max}$. The grey dashed line in the $v_\perp$ vs. $v_{\parallel}$ plot indicates the line of equality, i.e.\ a spherically symmetric velocity surface. Computing the cross-sectional velocity requires an estimate of the distance to the host, where we here assume $D_L = 44.2 \pm 2.3$\,Mpc derived from the Planck cosmology \citep{Planck2018} and the cosmological recession velocity of the host galaxy, $z_{\rm cosmic} = 0.00986 \pm 0.00049$ \citep{Mukherjee2021,Sneppen2023_H0}. We do not propagate the 5\% uncertainty in the absolute luminosity distance into $v_\perp$, as we are interested in the relative change and uncertainties in fitting the spectra themselves. In the right panels, we illustrate the evolution in the continuum (top) and in the 1\,$\mu$m absorption feature (bottom) - which produces the subtle difference in the best-fit parameters. Additional systematic effects like line-blending could shift the posteriors, but are likely to affect sub-epochs in a similar fashion. } 
    \label{fig:Epoch1_param}
\end{figure*}

\section{Observed sub-epoch evolution}\label{sec:delta}
In Fig.~\ref{fig:Epoch 1}, we show the epoch~1 spectra from the nodding-mode reduction, which clearly shows that the transient is subtly, but statistically significantly, evolving between the exposures. The drop in flux is particularly pronounced near the blackbody peak and in the UV, while the two spectra converge towards the redder wavelengths of the optical. 

To constrain and quantify the spectral evolution, we model the spectra as a blackbody perturbed with a P~Cygni profile from the three strong lines of \srii at 1.0037, 1.0327, and 1.0915\,\(\mu\)m following the framework detailed in \citet{Watson2019} and \citet{Sneppen2023,Sneppen2023_H0}. The flux drop close to the blue edge of the spectrograph at around 350--400\,nm is likely due to \yii and \zrii absorption \citep{Gillanders2022,Shingles2023,Sneppen2023_yttrium,Vieira2023}. Hence the spectral shape in this region of the spectrum is degenerate with the modelling of the UV line-opacity. Therefore we only fit the wavelengths $\lambda\geq400\,$nm, where the spectrum is better constrained. In the following, we consider the evolution of the continuum and spectral features.

We quickly summarise the parameters investigated. First, the blackbody continuum is described by two parameters, the blackbody temperature, $T_{\rm obs}$ and velocity/radius, $v_{\perp}= R_{ph}/t$, which respectively decide the location of the spectral peak and the normalisation of the spectrum. We note that computing the velocity from the normalisation (ie. angular size) requires an assumption of cosmological distance. Here we assume for NGC\,4993 is $44.2 \pm 2.3$\,Mpc (derived from the cosmic recession velocity of $z_{\rm cosmic} = 0.00986 \pm 0.00049$; \citealt{Mukherjee2021,Sneppen2023_H0}, and assuming Planck cosmology; \citealt{Planck2018}), but any typically prescribed distance yields similar results. Second, the 0.7-1.0\,$\mu$m feature is modelled as a P~Cygni profile from the strong three lines of \srii (4p\(^6\)4d---4p\(^6\)5p) as detailed in \cite{Watson2019}. The line profile is parameterised using the spherically symmetric Elementary Supernova model in \cite{Jeffery1990} with a line optical depth $\tau$, a velocity stratification with a scaling velocity $v_{\rm e}$, a photospheric velocity of the line $v_{\rm \parallel}$, and a maximum ejecta velocity $v_{\rm max}$. Further discussion on comparing the inner line-forming region (ie. $v_{\rm \parallel}$) with the emitting area of the blackbody (ie. $v_{\rm \perp}$) can be found in \cite{Sneppen2023}. 

Notably, this P~Cygni framework assumes a pure scattering regime, where any photons absorbed will be re-emitted through the same transition (i.e. without fluorescing into other lines). In the \srii case, this means that after absorption, the upper energy-level, 4p\(^6\)5p, will decay down to the lower-level of the transitions, 4p\(^6\)4d, and not the ground-state (4p\(^6\)5s). This is a good approximation, as the Sobolev optical depth for the ground state transition is 1-2 orders-of-magnitude larger and thus has a much smaller Sobolev escape probability. Conversely, excess emission (relative to the absorption) could potentially be produced from fluorescence from the ground-state (ie. an excitation from 4p\(^6\)5s to 4p\(^6\)5p, which then decays to 4p\(^6\)4d). Such a systematic effect would make it more tenuous to draw tight physical constraints (such as the electron density discussed in Sec. \ref{sec:emergence}) from the observed emission/absorption ratio. However, this will likely not strongly affect the temporal formation of the feature and indeed observationally the emission is not boosted but instead largely absent in the 1.4\,day spectra.

\subsection{Continuum evolution}\label{sec:3.1_continuum}

\begin{figure}    \includegraphics[width=\linewidth,viewport=12 20 405 320, clip=]{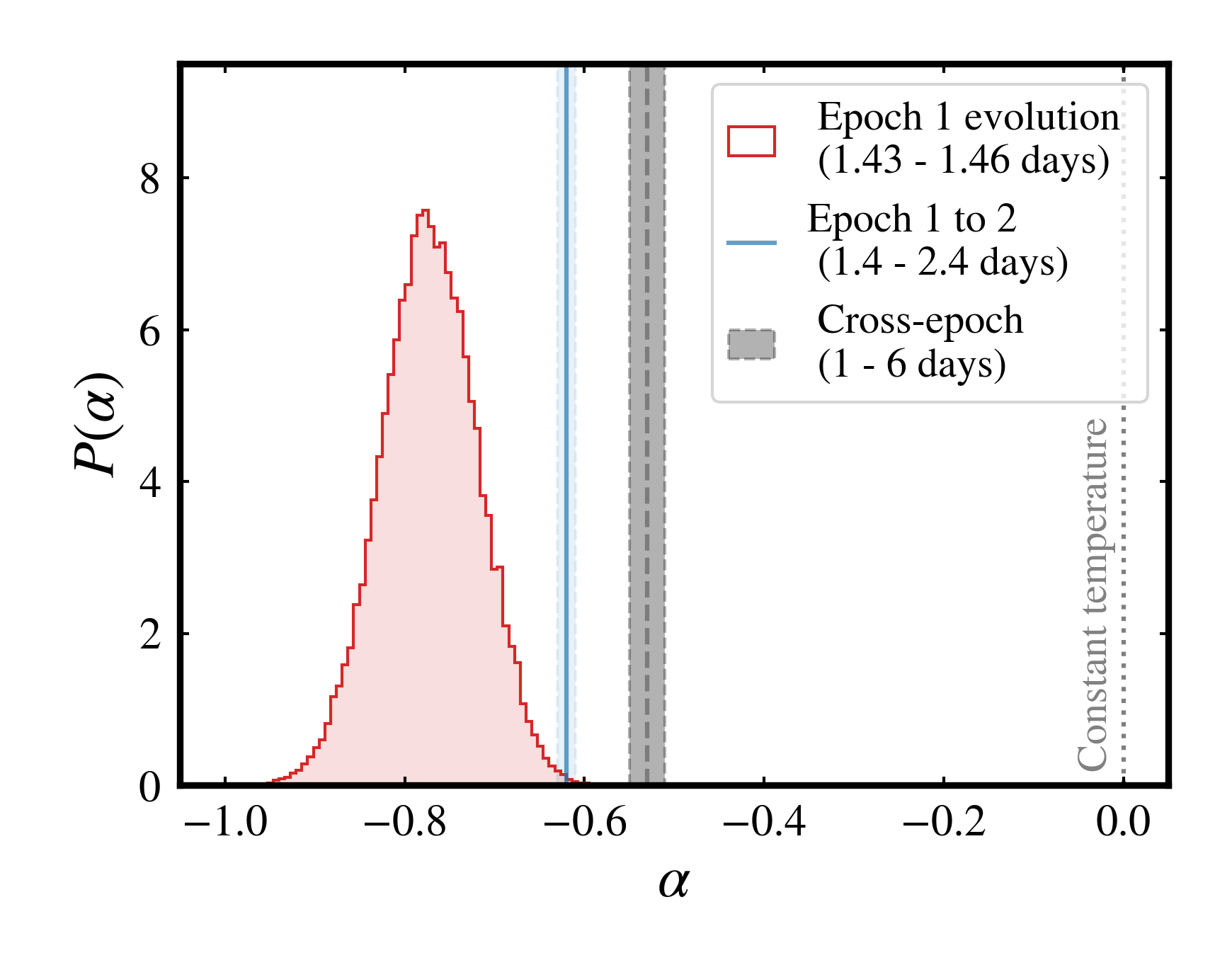}
    \caption{The sub-epoch evolution of the power-law index of cooling (see Eq.~\ref{eq:1}) of the X-shooter spectra at 1.43 days post-merger (red histogram). The temperature evolution is highly significant ($>10\sigma$ from $\alpha=0$). The inferred cooling between exposures, $\alpha=-0.77 \pm 0.06$, is more rapid than inferred from the average epoch~1--2 evolution ($\alpha=-0.62 \pm 0.01$, \citealt{Sneppen2023}), which is itself more rapid than the cross-epoch analysis over the period 1--6 days post-merger \citep{Waxman2018}.} 
    \label{fig:alpha}
\end{figure}

In Fig.~\ref{fig:Epoch 1}, lower panel, we plot the ratio of fluxes for different wavelength bins. The wavelength-dependent decline in flux predicted from a blackbody cooling in time (indicated with grey shading) traces the UV and optical decline fairly well. We emphasise this is not a fit to the data, but a prediction of a blackbody with a powerlaw decline of temperature across time derived from comparing epochs 1--2 (see Sec. \ref{sec:predicted}). On these short timescales the evolution in flux from 400\,nm to 1000\,nm broadly follows the predictions of blackbody cooling. Fitting the blackbody temperature and normalisation for each spectrum independently highlights the evolution in ejecta parameters (see Fig.~\ref{fig:Epoch1_param}). The best-fit temperature clearly cools between the exposures and the emitting area inferred from the blackbody normalisation suggests that $v_{\perp}$ has decreased. 
  
The constraints on the temperature evolution are further quantified in Fig.~\ref{fig:alpha}. The inferred $\alpha=-0.77 \pm 0.06$ from the sub-epoch evolution suggests a rapid cooling, which is faster than the 1--2 epoch evolution ($\alpha=-0.62 \pm 0.01$, \citealt{Sneppen2023}) and the average 1--6 days post-merger evolution ($\alpha=-0.54 \pm 0.02$, \citealt{Waxman2018}). This suggests that the early temperature evolution is steeper than inferred from comparing with later epochs. Naturally, the statistical constraining power is lower due to the smaller temporal range, but it is precisely this temporal locality which makes this estimate complementary in nature to the cross-epoch (but temporally sparse) analysis. That the powerlaw-slope, $\alpha$, varies with time is a strong indication that a single powerlaw prescription for the temperature, while a useful and common parameterization, is an oversimplification. 

That $v_\perp$ decreases may indicate we are observationally probing the recession of the thermalisation surface, which drops deeper into the ejecta as the outer layers become increasingly optically thin. While it is possible that the variable seeing with non-optimal modelling of the slitloss could produce a similar shift, nevertheless, we believe the effect we are observing is real, as there is no other indication of strong slitloss effects and, as we discuss in Sect.~\ref{sec:3.2}, the \srii line also suggests a recession of the photosphere. 


\subsection{Spectral features}\label{sec:3.2}

The \(1\,\mu\)m line has a P~Cygni nature, which allows constraints on the velocity of the line-forming region. We show these constraints as part of the model fit in 
Fig.~\ref{fig:Epoch1_param}. The line photospheric velocity, $v_{\parallel}$, and the outer velocity of the line-forming region, $v_{\rm max}$, decrease slightly with time. This independently illustrates and constrains the photospheric surface recession into the ejecta. We note here that, similar to the expanding photosphere method for AT2017gfo presented in \cite{Sneppen2023}, Fig.~\ref{fig:Epoch1_param} clearly shows that the thermalisation radius of the blackbody and the photospheric radius of the line decrease coherently and remain consistent with each other across the exposures within the first epoch. The constraints presented in Fig.~\ref{fig:Epoch1_param}, only include the statistical uncertainty of the fitting model; systematic effects, such as line-blending, time-delay or reverberation-effects for $v_{\parallel}$ or the uncertainty in luminosity distance for $v_{\perp}$, could potentially shift the means of the posteriors, but such effects will likely affect sub-epochs in a similar way, so the temporal evolution of the fitted parameters still follows the trend suggested from cross-epoch analysis. 

As mentioned above, the absorption feature below 400\,nm (the UV deficit compared to the blackbody) in the first epochs has been interpreted as tentative evidence of \yii and \zrii lines \citep{Gillanders2022,Vieira2023}. While we do not fit these ranges in the model, we can still see that the change in flux over these wavelengths is more pronounced between exposures and may indicate increasing opacity at short wavelengths and thus more reprocessing of the light towards redder colours. By later epochs, this absorption feature has disappeared, which may follow from lanthanide (or d-block element) line-blanketing washing out all spectral features below 600--700\,nm  \citep{Gillanders2022,Sneppen2023_yttrium}.

The NIR spectra are the most difficult to interpret, partly due to the smaller SNR, partly because the atmospheric conditions can also change on these timescales (see Sec \ref{sec:atmospheric}) and
partly because some of the strongest emission features appear in the NIR at later epochs, so contributions from such features may also be present and evolving at this time. Such early NIR emission features are still poorly understood, but must be relatively minor at this epoch given the observed continuum.
Regardless, the decrease in the bulk flux is least in the NIR, and seems to be more or less as predicted from the cooling blackbody framework.




\begin{figure}
\includegraphics[width=\columnwidth,viewport=18 20 490 380, clip=]{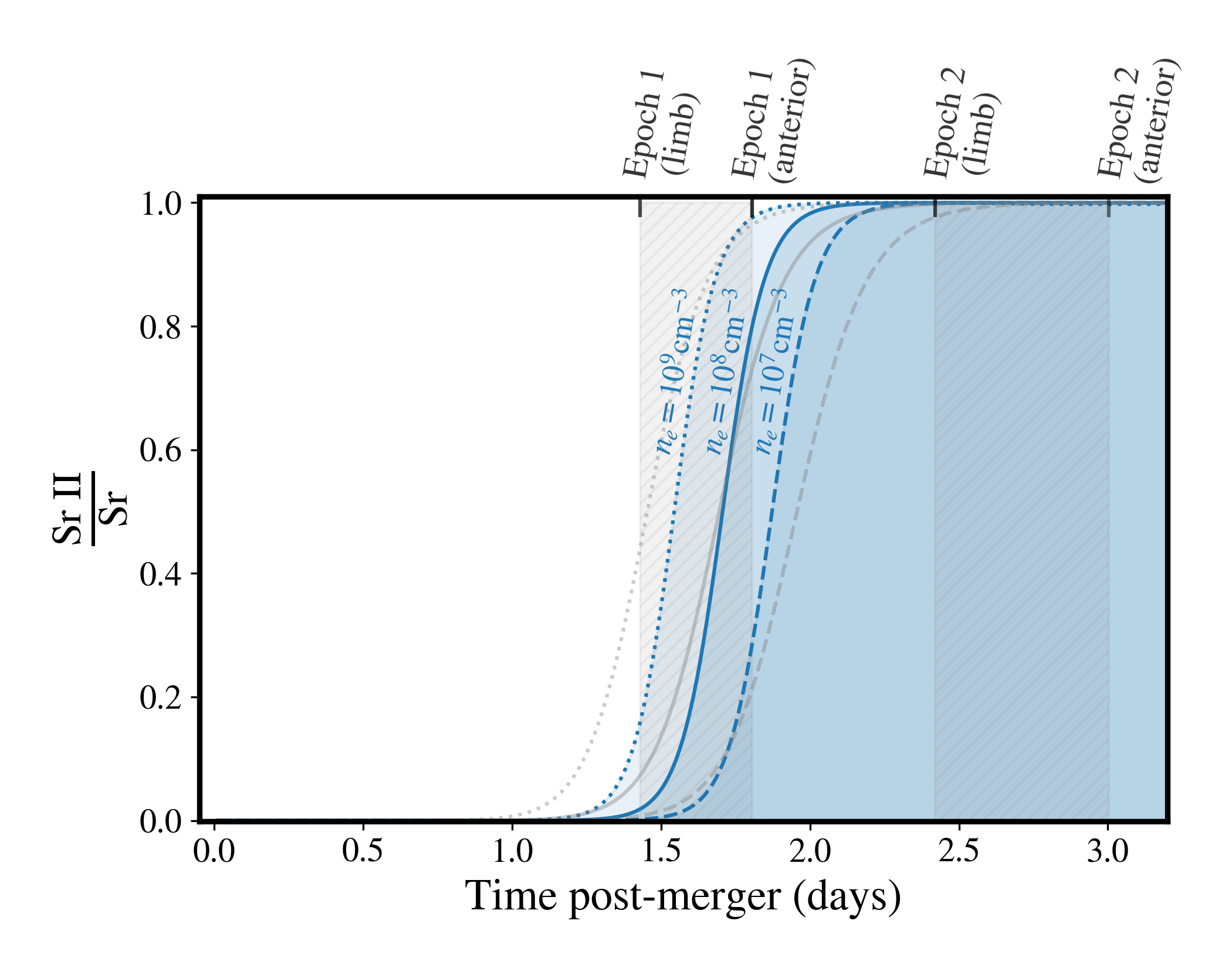}
    \caption{Fraction of Sr in its first ionised state given the time since the merger in LTE. The \srii is shown as a fraction of total Sr and is transitioning (recombining) from the \sriii state. The blue lines are based on the temperature evolution from the sub-epoch analysis in this paper, while the grey background lines assume the slower cooling in \citet{Waxman2018}. Dashed and dotted lines show the fraction assuming different electron densities. Regardless of the exact cooling-rate, and across a broad range of electron densities, there is a sharp transition from \sriii to \srii at 1.4--1.7 days post-merger (or equivalently for an observer the absorption will form at 1.0--1.2 days). This suggests that 1) spectra taken just a few hours earlier would have a significantly different \srii profile and 2) the more distant ejecta will have less \srii. On the upper axis we show the time of the X-shooter epochs 1--2 \citep{Pian2017,Smartt2017}. We note that due to light-travel time delays the more distant ejecta (near the limb) is observed at a different time to the anterior ejecta as indicated with the grey shaded regions.} 
    \label{fig:sr_appearance}
\end{figure}

\section{Discussion}\label{sec:discussion}
In this manuscript, we have reduced and compared the sub-epochs of the first phase of observations obtained with X-shooter for AT2017gfo. We have highlighted the physical information of the kilonova these changes on short-timescale contain, complementary to the sparse long-time scales between observations. These considerations naturally lead to a discussion on the constraints attainable given observations with better cadence. In Sect.~\ref{sec:emergence} below, we therefore discuss the additional constraints for line-identifications a higher cadence allows and in Sect.~\ref{sec:cadence} we discuss how better to optimise future observing strategies.

\begin{figure*}
    \centering
    \includegraphics[width=\linewidth,viewport=80 10 990 480, clip=]{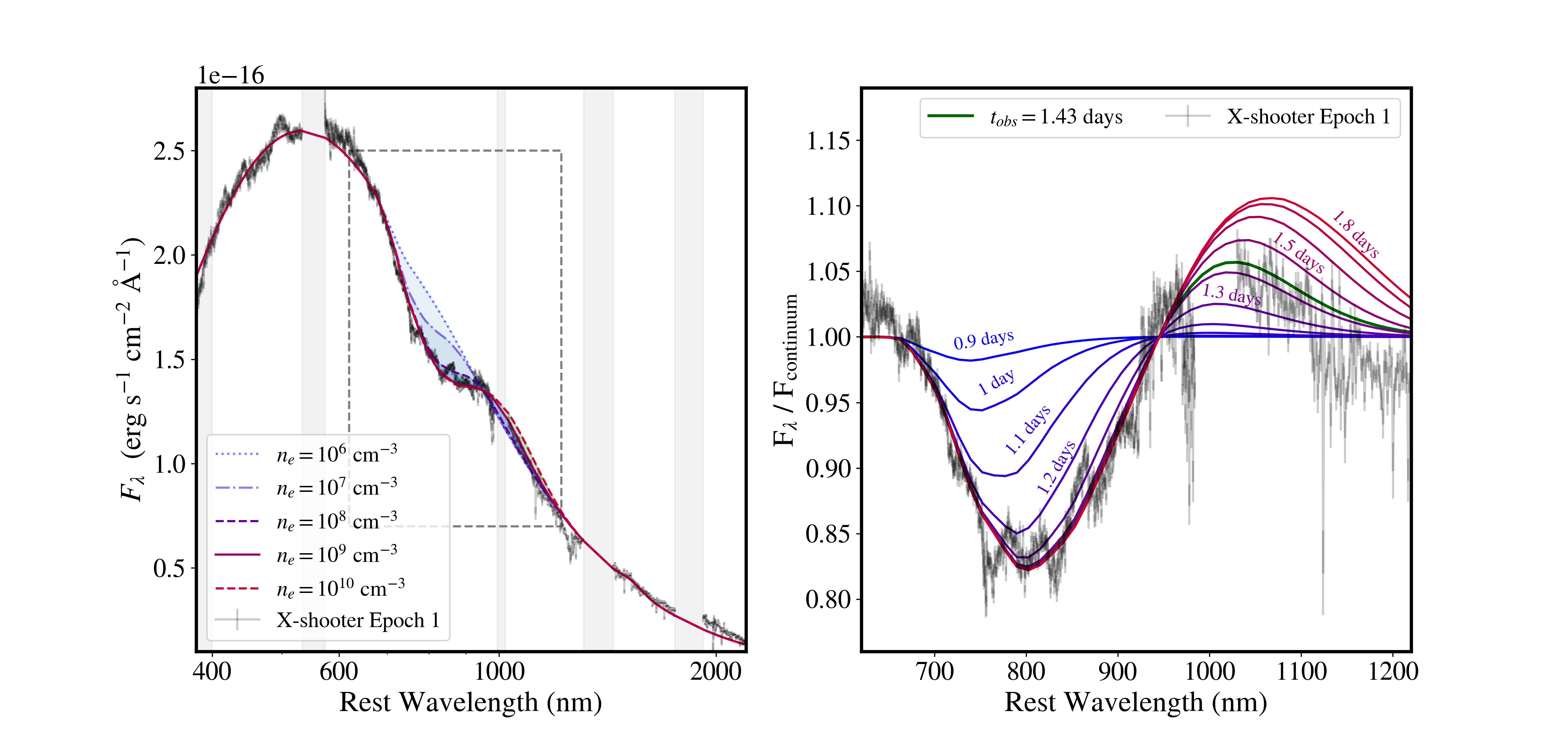}
    \caption{ The emergence of a \srii P~Cygni given different electron densities (left) and time post-merger (right). \emph{Left:} The X-shooter epoch~1 spectrum with a model blackbody + \srii  P~Cygni feature overlaid. The wavelength-dependent strength of the \srii feature is modulated by the fractional abundance of single ionised Sr at the time of observation, $t_{\rm obs}=1.43$ days (see Fig.~\ref{fig:sr_appearance}). There is only a weak dependence on the electron density as the transition-time is a relatively robust prediction in time. This means that the electron density is not tightly constrained, with the majority of the feature indicating a moderate-to-high electron density of $n_e \approx 10^8-10^{9}$ cm$^{-3}$. 
    \emph{Right:} given an electron density $n_e = 10^9$ cm$^{-3}$, we show the corresponding \srii P~Cygni from 0.9 to 1.8 days post-merger in intervals of 0.1 days. The feature rapidly emerges with the reverberation wave moving from the blueshifted (and nearer) ejecta towards the red. At the time of observation the absorption is fully formed while the emission is weak due to the decreased fraction of \srii in the emitting region at the time of emission. This means that the \srii interpretation makes strong predictions for the timing of the emergence of the feature and the early spectral shape and its evolution. } 
    \label{fig:PCygni_Emergence}
\end{figure*}


\begin{figure}
    \centering
    \includegraphics[width=\linewidth,viewport=15 20 450 450, clip=]{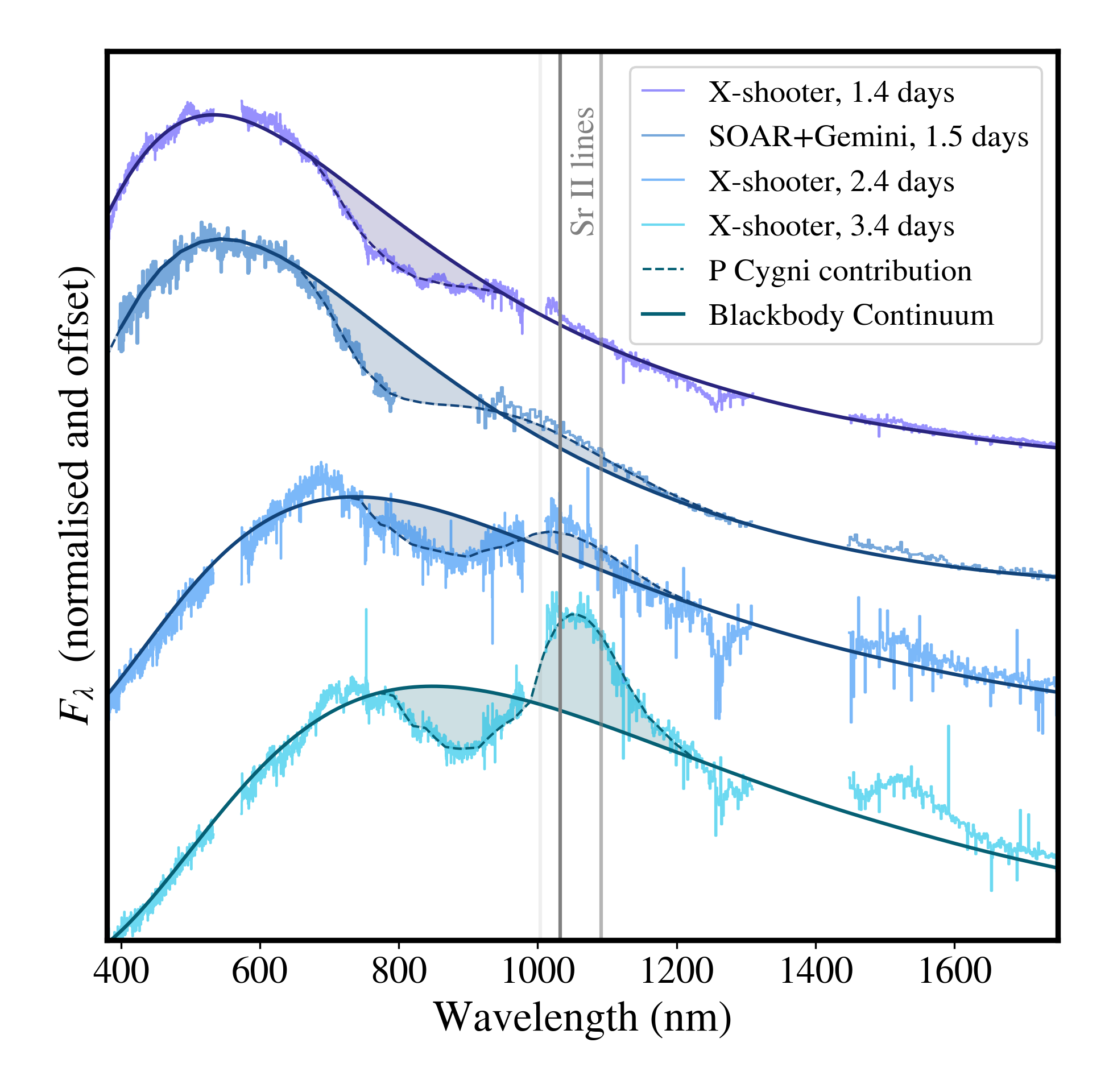}
    \caption{ X-shooter epochs 1-3 (from 1.4-3.4 days post-merger) alongside SOAR \citep{Nicholl2017} and Gemini-South \citep{Chornock2017} spectra from 1.5 days. As predicted from the 1.4 day blackbody-temperature the recombination-wave passing through the ejecta has not reached the equatorial ejecta, resulting in no emission peak. However, just hours later and in the subsequent days \sriii has recombined to \srii throughout the ejecta forming both a blueshifted absorption valley and an emission peak - in essence a full P Cygni profile. Modelling the continuum as a simple blackbody becomes increasingly tenuous after the first two days - for example blueward of the 1\,$\mu$m feature, around 700\,nm the \yii 4d\(^2\)--4d5p transitions are expected to perturb the SED \citep{Sneppen2023_yttrium}, while UV line-blanketing are likely observationally needed to match the spectral shape below $\lambda\lessapprox$700\,nm \citep{Gillanders2022}.  }
    \label{fig:14_24day_spectrum}
\end{figure}
 
\subsection{The rapid emergence of spectral lines}\label{sec:emergence}
Conventionally, line identifications are determined by matching observed features with modelled spectral lines which have consistent transition wavelength and transition probabilities. Thus, identifications follow from analysing the wavelength dimension of the data, while spectra at different times merely serve as different spectral samples for wavelength models. However, for transients with rapid evolution, the timing of the emergence of spectral features is itself a strong prediction for specific line-identifications both in terms of which ionisation states are dominant and in terms of being able to deblend lines, as we exemplify with two of the proposed identifications below. 

\paragraph{\srii:} With a second ionisation energy for Sr of 11.0\,eV, typical KN electron densities (i.e.\ $n_e \approx 10^7-10^9$\,cm$^{-3}$) and LTE conditions, \sriii (or even higher-ionised species) will become the dominant species for temperatures greater than $T \approx 4300-4700$\,K. For comparison, the emitted blackbody temperature for the ejecta nearest the observer in epoch~1, $T(\cos(\theta)=1,t_{\rm obs} = 1.43 \ {\rm days}) = 4150 \pm 60$\,K, while the ejecta further from the observer (which due to light travel-time is observed at an earlier and hotter time) $T(\cos(\theta)=\beta,t_{\rm obs} = 1.43 \ {\rm days}) = 4900 \pm 70$\,K \citep{Sneppen2023_bb}. Here, $\theta$ is the angle between the direction of expansion and the line-of-sight. The transition between ionisation states is closely dictated by the temperature because of its exponential dependency in LTE, while the transition time is much less sensitive to the exact electron density. 
As the ionisation-state transitions, the optical depth, $\tau$, of the line will change by orders of magnitude over a few hours. In the radiative transfer equations, this implies a drastic change in the transmission, $I = I_0 e^{-\tau}$, and thus a rapid formation of the line. This immediately suggests two temporal predictions of the \srii appearance. 
    
First, spectra taken significantly earlier than about 1.2 day post-merger should not have a \(1\,\mu\)m P~Cygni feature due to \srii (as shown in Fig.~\ref{fig:sr_appearance}). This fraction of \srii is computed assuming LTE with the Saha ionisation equation and the temporal dependence on temperature in Eq.~\ref{eq:1} with the cooling-rate, $\alpha\approx0.77$, inferred in Sect.~\ref{sec:3.1_continuum}. Applying a slower cooling-rate such as that suggested in \citet{Waxman2018} does not significantly change the rapid nature of the recombination\footnote{Or indeed, \emph{combination}, as it may be the first time these freshly-formed atoms have ever reached this state.} and the emergence time is only relatively weakly dependent on the exact electron density. In Fig.~\ref{fig:PCygni_Emergence} (right panel), we illustrate how rapidly a \srii feature would emerge.  

Second, the early spectral shape of \srii is very distinct due to light travel-time effects. The more distant (earlier and thus hotter) parts of the ejecta will have more \sriii and less \srii (Fig.~\ref{fig:sr_appearance}). As the distant ejecta is the origin of the emission peak of the P~Cygni, and the nearer parts are the origin of the absorption, this would suggest less emission relative to absorption. That is, at early times a \srii feature would emerge in absorption blueshifted away from the rest-wavelength of the line, but at increasingly later times as the recombination wave passes through the ejecta, a full P~Cygni feature would emerge. This seems to be observationally suggested by the spectra of epoch~1, where the absorption dominates and the emission peak is small (see Fig.~\ref{fig:PCygni_Emergence}) and the later epochs where the emission feature grows in prominence (see Fig. \ref{fig:14_24day_spectrum}). Indeed the spectra from Gemini-South, taken just an hour after X-shooter, displays a subtle emission peak, formed within the narrow time-frame as predicted for a Strontium recombination-wave.
    
In general, given a line-ID (i.e.\ assuming an ionisation-energy) the timing of the feature emergence allows the electron density to be determined under LTE conditions. Singly- to doubly-ionised \rprocess ejecta with $M_{\rm ejecta} = 0.05 M_{\odot}$, $v=0.25c$ (KN characteristic scales from e.g. \cite{Kasen2017}) and $t=1.43$ days would have a mean electron density: 
$\log_{10}(n_e/{\rm cm}^{-3}) = 8.3-8.6$. Naturally, many systematic effects could shift this estimate substantially, but this density is notably close to that inferred from the 1$\mu$m feature given the Sr interpretation, $\log_{10}(n_e/{\rm cm}^{-3}) \approx 8-9$ (see Fig.~\ref{fig:PCygni_Emergence}, left panel). Conversely, assuming a broad range of electron densities one could determine what ionisation energy would be compatible with the line-shape and the formation time (i.e.\ putting strong constraints on the spectral line-ID). We emphasise, that \srii (under typical KN electron densities) does match the observed feature's formation provides a strong observational indication that the blackbody radiation temperature sets the electron temperature of the ejecta. The optimal constraints on this recombination wave would be a series of spectra detailing the feature shift over several hours (see Fig.~\ref{fig:PCygni_Emergence}, right panel). 
    
Earlier spectra than the VLT/X-shooter epoch~1 were taken with Magellan telescopes using the LDSS-3 and MagE spectrographs (at 0.49 and 0.53 days post-merger) and indeed show no \(0.7-1\,\mu\)m feature \citep{Shappee2017}. A later spectrum taken using the ANU 2.3m, WiFeS instrument \citep[0.93 days,][]{Andreoni2017} displays only weak or non-existent absorption, while the South African Large Telescope spectrum \citep[1.18 days,][]{Buckley2018} has a sizeable absorption component as predicted from Fig.~\ref{fig:PCygni_Emergence}. However, spectral modelling is difficult as these spectra have relatively low S/N and only X-shooter (with its UV-NIR coverage) provides strong constraints on the continuum shape and the \(1\,\mu\)m emission peak. While this emergence time is thus promising for the \srii interpretation, we leave a detailed comparative analysis for future work (Sneppen et~al., in preparation). It remains to be seen whether a \ion{He}{i} interpretation of the $1\,\mu$m feature, as introduced in \citet{Perego2022} and argued for in \citet{Tarumi2023}, can reproduce both the observed rapid absorption-to-emission evolution and the time of appearance, both of which seem to be required by the \ion{Sr}{ii} identification.

At late times, when the \srii species becomes depopulated the feature will recede and again due to reverberation this will first be in the blueshifted ejecta (i.e.\ leaving an emission peak which then subsequently fades away). The modelling of the reverberation discussed here mainly focuses on the changing ionisation. The change in the strength of the observed lines and the fading of the spectral continuum with time could produce additional reverberation effects as discussed in \cite{Sneppen2023_H0}. However, modelling the changing continuum only produces a minor effect at early times because of the softer flux evolution towards the Rayleigh-Jeans--tail of the blackbody. Nevertheless, these continuum reverberation effects are essential for understanding the \(0.7-1\,\mu\)m feature for intermediate-to-late time evolution when the continuum shifts redward and the feature's emission peak is observed to become dominant (see McNeill et~al.\, in~preparation). 

This reverberation of line-formation has primarily focused on ionisation-states and not the temporal change of the level-populations. We have assumed LTE level populations, and while this means that only 3-7\% of \srii present will be excited to the 4p\(^6\)4d levels conducive to producing the observed NIR triplet absorption feature (for T in the range $4150-4900$\,K), the high Sr mass fraction for a solar \rprocess abundance ratio \citep{Lodders2009,Bisterzo2014} and the intrinsically very strong transitions means this feature is still expected to be prominent. Using the Boltzmann formula given the relevant temperatures, 4150-4900 K, one can compute the LTE level population. This only has a weak dependence on temperature relative to the ionisation-ratio, for instance when 4p\(^6\)4d population (the relevant lower level for the transition) changes by a factor of 2 relative to the \srii ground state, then the \sriii/\srii ratio will change by two orders of magnitude. 

Lastly, we emphasise that non-LTE effects, a highly in-homogeneous temperature distribution (and potentially a complex radial/angular density distribution) could make the transitions between ionisation states more gradual. Full radiative transfer modelling of merger simulations is required to explore these dependencies. Conversely, strong observational constraints on the emergence of lines can thus potentially reveal insights into the homogeneity of temperature and the validity of LTE-approximations in the ejecta. 
    
    
\paragraph{\yii:} There are two series of lines from the 4d\(^2\)--4d5p transitions of \yii, one prominent set of lines centred around 760\,nm and slightly weaker transitions at 670\,nm \citep{Biemont2011}. For the early characteristic expansion velocities (i.e.\ $v\approx 0.3c$) the features from these components will blend together, where the emission and absorption features will produce a series of complex wiggles in the spectra - particularly impacting the SED around 700\,nm. However, as velocities decrease these lines gradually deblend. When the characteristic velocity recedes below $0.2c$ (around 4 days post-merger) a P~Cygni feature should emerge from the 760\,nm lines if yttrium is responsible for these lines. We emphasise that the time when the feature should first appear is a prediction specific to the configuration of atomic levels in \yii. This feature is indeed found to emerge clearly between epochs 3--4 for AT2017gfo \citep{Sneppen2023_yttrium}.     


\subsection{Optimal observing strategy}\label{sec:cadence}
While the timescale of evolution for astrophysical objects, even most transients, is typically much longer than the exposure time, for fast-evolving transients like KNe this is not always the case. The situation is likely to be even more extreme for earlier identifications post-merger than for AT2017gfo. Furthermore, given the emergence of \emph{JWST} as a viable tool to produce spectra of gamma-ray burst--triggered KNe \citep{Rastinejad2022,Levan2023} as well as the improvements to the LIGO sensitivity and thus the increased distance of most future KNe, reaching a similar signal-to-noise ratio (S/N) as obtained for AT2017gfo will likely require longer exposure times, at least before the advent of the next generation of extremely large ground-based optical telescopes. Thus, in the context of the physical information contained within this short timescale perspective, it is natural to consider what would be an optimal observing strategy for future KNe. 

First, higher cadence can offer remarkable constraining power for KN modelling --- especially at specific temporal intervals (such as around 1.0-1.5 days post-merger) where sharp transition in ionisation-states occurs. As the temperature and density evolution is rapid, species cross the threshold between their ionisation states on the timescale of an hour (i.e.\ their corresponding spectral lines will emerge or disappear rapidly), the photosphere recedes deeper within the ejecta, and lines deblend to form interpretable features (see Sect.~\ref{sec:emergence}). Thus, even if a future KN due to observational limitations is only briefly observable (such as two consecutive hours) it is highly constraining to leverage the full temporal-span to study the evolutionary trend in parameters. 

Second, observing blocks should prioritise several shorter exposures rather than few long exposures, particularly within the first day or two post-merger. The overhead associated with increased readout time is minor compared to the gain in information from probing the KN evolution between individual exposures. For instance in the case of AT2017gfo, the four exposures in epoch~1 allow for both temporally-resolved nodding and stare-mode reductions. Conversely, the two (albeit longer) exposures taken in epoch~2 limits the temporal information and removes the possibility of a robustness-test of the reduction pipeline across sub-epochs. Ideally, the observing dither pattern would allow a series of individual nodding mode reductions (e.g. in a repeating ABBA dither pattern). 

Third, while spectra derived from averaging several exposures for optimal S/N are the standard, early sub-epoch spectra contain additional insights into the time derivative of KN properties and therefore should ideally be reduced and published as well, to provide the full synergy of the time-series spectral evolution.

\section*{Data availability}
Work in this paper was based on observations made with European Space Observatory (ESO) telescopes at the Paranal Observatory under programmes 099.D-0382 (principal investigator E. Pian), 099.D-0622 (principal investigator P. D’Avanzo), 099.D-0376 (principal investigator S. J. Smartt). The data are available at http://archive.eso.org. The re-reduced sub-epoch spectra are made available from: \url{https://github.com/Sneppen/Kilonova-analysis}

\section*{Acknowledgements}
We thank Jonatan Selsing for useful correspondence on the original reduction pipeline in \cite{Pian2017} and Tom Reynolds for the implementation of \texttt{Molecfit}. We thank Stephen Smartt, Stuart Sim, Christine Collins and Luke Shingles for comments and discussions on the evolution of the recombination-wave. The Cosmic Dawn Center (DAWN) is funded by the Danish National Research Foundation under grant DNRF140. AS and DW are funded by the European Union (ERC, HEAVYMETAL, 101071865). Views and opinions expressed are however those of the authors only and do not necessarily reflect those of the European Union or the European Research Council. Neither the European Union nor the granting authority can be held responsible for them.

\bibliographystyle{mnras}
\bibliography{refs} 

\begin{thebibliography}{}
\makeatletter
\relax
\def\mn@urlcharsother{\let\do\@makeother \do\$\do\&\do\#\do\^\do\_\do\%\do\~}
\def\mn@doi{\begingroup\mn@urlcharsother \@ifnextchar [ {\mn@doi@} {\mn@doi@[]}}
\def\mn@doi@[#1]#2{\def\@tempa{#1}\ifx\@tempa\@empty \href {http://dx.doi.org/#2} {doi:#2}\else \href {http://dx.doi.org/#2} {#1}\fi \endgroup}
\def\mn@eprint#1#2{\mn@eprint@#1:#2::\@nil}
\def\mn@eprint@arXiv#1{\href {http://arxiv.org/abs/#1} {{\tt arXiv:#1}}}
\def\mn@eprint@dblp#1{\href {http://dblp.uni-trier.de/rec/bibtex/#1.xml} {dblp:#1}}
\def\mn@eprint@#1:#2:#3:#4\@nil{\def\@tempa {#1}\def\@tempb {#2}\def\@tempc {#3}\ifx \@tempc \@empty \let \@tempc \@tempb \let \@tempb \@tempa \fi \ifx \@tempb \@empty \def\@tempb {arXiv}\fi \@ifundefined {mn@eprint@\@tempb}{\@tempb:\@tempc}{\expandafter \expandafter \csname mn@eprint@\@tempb\endcsname \expandafter{\@tempc}}}

\bibitem[\protect\citeauthoryear{{Andreoni} et~al.,}{{Andreoni} et~al.}{2017}]{Andreoni2017}
{Andreoni} I.,  et~al., 2017, \mn@doi [\pasa] {10.1017/pasa.2017.65}, \href {https://ui.adsabs.harvard.edu/abs/2017PASA...34...69A} {34, e069}

\bibitem[\protect\citeauthoryear{{Bi{\'e}mont} et~al.,}{{Bi{\'e}mont} et~al.}{2011}]{Biemont2011}
{Bi{\'e}mont} {\'E}.,  et~al., 2011, \mn@doi [\mnras] {10.1111/j.1365-2966.2011.18637.x}, \href {https://ui.adsabs.harvard.edu/abs/2011MNRAS.414.3350B} {414, 3350}

\bibitem[\protect\citeauthoryear{{Bisterzo}, {Travaglio}, {Gallino}, {Wiescher}  \& {K{\"a}ppeler}}{{Bisterzo} et~al.}{2014}]{Bisterzo2014}
{Bisterzo} S.,  {Travaglio} C.,  {Gallino} R.,  {Wiescher} M.,   {K{\"a}ppeler} F.,  2014, \mn@doi [\apj] {10.1088/0004-637X/787/1/10}, \href {https://ui.adsabs.harvard.edu/abs/2014ApJ...787...10B} {787, 10}

\bibitem[\protect\citeauthoryear{{Buckley} et~al.,}{{Buckley} et~al.}{2018}]{Buckley2018}
{Buckley} D. A.~H.,  et~al., 2018, \mn@doi [\mnras] {10.1093/mnrasl/slx196}, \href {https://ui.adsabs.harvard.edu/abs/2018MNRAS.474L..71B} {474, L71}

\bibitem[\protect\citeauthoryear{{Chornock} et~al.,}{{Chornock} et~al.}{2017}]{Chornock2017}
{Chornock} R.,  et~al., 2017, \mn@doi [\apjl] {10.3847/2041-8213/aa905c}, \href {https://ui.adsabs.harvard.edu/abs/2017ApJ...848L..19C} {848, L19}

\bibitem[\protect\citeauthoryear{{Collins} et~al.,}{{Collins} et~al.}{2023}]{Collins2023b}
{Collins} C.~E.,  et~al., 2023, \mn@doi [arXiv e-prints] {10.48550/arXiv.2309.05579}, \href {https://ui.adsabs.harvard.edu/abs/2023arXiv230905579C} {p. arXiv:2309.05579}

\bibitem[\protect\citeauthoryear{{Coulter} et~al.,}{{Coulter} et~al.}{2017}]{Coulter2017}
{Coulter} D.~A.,  et~al., 2017, \mn@doi [Science] {10.1126/science.aap9811}, \href {https://ui.adsabs.harvard.edu/abs/2017Sci...358.1556C} {358, 1556}

\bibitem[\protect\citeauthoryear{{Cowperthwaite} et~al.,}{{Cowperthwaite} et~al.}{2017}]{Cowperthwaite2017}
{Cowperthwaite} P.~S.,  et~al., 2017, \mn@doi [\apjl] {10.3847/2041-8213/aa8fc7}, \href {https://ui.adsabs.harvard.edu/abs/2017ApJ...848L..17C} {848, L17}

\bibitem[\protect\citeauthoryear{{Domoto}, {Tanaka}, {Kato}, {Kawaguchi}, {Hotokezaka}  \& {Wanajo}}{{Domoto} et~al.}{2022}]{Domoto2022}
{Domoto} N.,  {Tanaka} M.,  {Kato} D.,  {Kawaguchi} K.,  {Hotokezaka} K.,   {Wanajo} S.,  2022, \mn@doi [\apj] {10.3847/1538-4357/ac8c36}, \href {https://ui.adsabs.harvard.edu/abs/2022ApJ...939....8D} {939, 8}

\bibitem[\protect\citeauthoryear{{Drout} et~al.,}{{Drout} et~al.}{2017}]{Drout2017}
{Drout} M.~R.,  et~al., 2017, \mn@doi [Science] {10.1126/science.aaq0049}, \href {https://ui.adsabs.harvard.edu/abs/2017Sci...358.1570D} {358, 1570}

\bibitem[\protect\citeauthoryear{{Fried}}{{Fried}}{1966}]{Fried1966}
{Fried} D.~L.,  1966, Journal of the Optical Society of America (1917-1983), \href {https://ui.adsabs.harvard.edu/abs/1966JOSA...56.1380F} {56, 1380}

\bibitem[\protect\citeauthoryear{{Gillanders}, {Smartt}, {Sim}, {Bauswein}  \& {Goriely}}{{Gillanders} et~al.}{2022}]{Gillanders2022}
{Gillanders} J.~H.,  {Smartt} S.~J.,  {Sim} S.~A.,  {Bauswein} A.,   {Goriely} S.,  2022, \mn@doi [\mnras] {10.1093/mnras/stac1258}, \href {https://ui.adsabs.harvard.edu/abs/2022MNRAS.515..631G} {515, 631}

\bibitem[\protect\citeauthoryear{{Jeffery} \& {Branch}}{{Jeffery} \& {Branch}}{1990}]{Jeffery1990}
{Jeffery} D.~J.,  {Branch} D.,  1990, in {Wheeler} J.~C.,  {Piran} T.,   {Weinberg} S.,  eds, ~{} Vol. 6, Supernovae, Jerusalem Winter School for Theoretical Physics. p.~149

\bibitem[\protect\citeauthoryear{{Kasen}, {Metzger}, {Barnes}, {Quataert}  \& {Ramirez-Ruiz}}{{Kasen} et~al.}{2017}]{Kasen2017}
{Kasen} D.,  {Metzger} B.,  {Barnes} J.,  {Quataert} E.,   {Ramirez-Ruiz} E.,  2017, \mn@doi [\nat] {10.1038/nature24453}, \href {https://ui.adsabs.harvard.edu/abs/2017Natur.551...80K} {551, 80}

\bibitem[\protect\citeauthoryear{{Kausch} et~al.,}{{Kausch} et~al.}{2015}]{Kausch2015}
{Kausch} W.,  et~al., 2015, \mn@doi [\aap] {10.1051/0004-6361/201423909}, \href {https://ui.adsabs.harvard.edu/abs/2015A&A...576A..78K} {576, A78}

\bibitem[\protect\citeauthoryear{{Levan} et~al.,}{{Levan} et~al.}{2023}]{Levan2023}
{Levan} A.,  et~al., 2023, \mn@doi [arXiv e-prints] {10.48550/arXiv.2307.02098}, \href {https://ui.adsabs.harvard.edu/abs/2023arXiv230702098L} {p. arXiv:2307.02098}

\bibitem[\protect\citeauthoryear{{Lodders}, {Palme}  \& {Gail}}{{Lodders} et~al.}{2009}]{Lodders2009}
{Lodders} K.,  {Palme} H.,   {Gail} H.~P.,  2009, \mn@doi [Landolt B\&ouml;rnstein] {10.1007/978-3-540-88055-4_34}, \href {https://ui.adsabs.harvard.edu/abs/2009LanB...4B..712L} {4B, 712}

\bibitem[\protect\citeauthoryear{{McCully} et~al.,}{{McCully} et~al.}{2017}]{McCully2017}
{McCully} C.,  et~al., 2017, \mn@doi [\apjl] {10.3847/2041-8213/aa9111}, \href {https://ui.adsabs.harvard.edu/abs/2017ApJ...848L..32M} {848, L32}

\bibitem[\protect\citeauthoryear{{Metzger} et~al.,}{{Metzger} et~al.}{2010}]{Metzger2010}
{Metzger} B.~D.,  et~al., 2010, \mn@doi [\mnras] {10.1111/j.1365-2966.2010.16864.x}, \href {https://ui.adsabs.harvard.edu/abs/2010MNRAS.406.2650M} {406, 2650}

\bibitem[\protect\citeauthoryear{{Mukherjee}, {Lavaux}, {Bouchet}, {Jasche}, {Wandelt}, {Nissanke}, {Leclercq}  \& {Hotokezaka}}{{Mukherjee} et~al.}{2021}]{Mukherjee2021}
{Mukherjee} S.,  {Lavaux} G.,  {Bouchet} F.~R.,  {Jasche} J.,  {Wandelt} B.~D.,  {Nissanke} S.,  {Leclercq} F.,   {Hotokezaka} K.,  2021, \mn@doi [\aap] {10.1051/0004-6361/201936724}, \href {https://ui.adsabs.harvard.edu/abs/2021A&A...646A..65M} {646, A65}

\bibitem[\protect\citeauthoryear{{Nicholl} et~al.,}{{Nicholl} et~al.}{2017}]{Nicholl2017}
{Nicholl} M.,  et~al., 2017, \mn@doi [\apjl] {10.3847/2041-8213/aa9029}, \href {https://ui.adsabs.harvard.edu/abs/2017ApJ...848L..18N} {848, L18}

\bibitem[\protect\citeauthoryear{{Perego} et~al.,}{{Perego} et~al.}{2022}]{Perego2022}
{Perego} A.,  et~al., 2022, \mn@doi [\apj] {10.3847/1538-4357/ac3751}, \href {https://ui.adsabs.harvard.edu/abs/2022ApJ...925...22P} {925, 22}

\bibitem[\protect\citeauthoryear{{Pian} et~al.,}{{Pian} et~al.}{2017}]{Pian2017}
{Pian} E.,  et~al., 2017, \mn@doi [\nat] {10.1038/nature24298}, \href {https://ui.adsabs.harvard.edu/abs/2017Natur.551...67P} {551, 67}

\bibitem[\protect\citeauthoryear{{Planck Collaboration} et~al.,}{{Planck Collaboration} et~al.}{2020}]{Planck2018}
{Planck Collaboration} et~al., 2020, \mn@doi [\aap] {10.1051/0004-6361/201833910}, \href {https://ui.adsabs.harvard.edu/abs/2020A&A...641A...6P} {641, A6}

\bibitem[\protect\citeauthoryear{{Rastinejad} et~al.,}{{Rastinejad} et~al.}{2022}]{Rastinejad2022}
{Rastinejad} J.~C.,  et~al., 2022, \mn@doi [\nat] {10.1038/s41586-022-05390-w}, \href {https://ui.adsabs.harvard.edu/abs/2022Natur.612..223R} {612, 223}

\bibitem[\protect\citeauthoryear{{Selsing} et~al.,}{{Selsing} et~al.}{2019}]{Selsing2019}
{Selsing} J.,  et~al., 2019, \mn@doi [\aap] {10.1051/0004-6361/201832835}, \href {https://ui.adsabs.harvard.edu/abs/2019A&A...623A..92S} {623, A92}

\bibitem[\protect\citeauthoryear{{Shappee} et~al.,}{{Shappee} et~al.}{2017}]{Shappee2017}
{Shappee} B.~J.,  et~al., 2017, \mn@doi [Science] {10.1126/science.aaq0186}, \href {https://ui.adsabs.harvard.edu/abs/2017Sci...358.1574S} {358, 1574}

\bibitem[\protect\citeauthoryear{{Shingles} et~al.,}{{Shingles} et~al.}{2023}]{Shingles2023}
{Shingles} L.~J.,  et~al., 2023, \mn@doi [\apjl] {10.3847/2041-8213/acf29a}, \href {https://ui.adsabs.harvard.edu/abs/2023ApJ...954L..41S} {954, L41}

\bibitem[\protect\citeauthoryear{{Smartt} et~al.,}{{Smartt} et~al.}{2017}]{Smartt2017}
{Smartt} S.~J.,  et~al., 2017, \mn@doi [\nat] {10.1038/nature24303}, \href {https://ui.adsabs.harvard.edu/abs/2017Natur.551...75S} {551, 75}

\bibitem[\protect\citeauthoryear{{Smette} et~al.,}{{Smette} et~al.}{2015}]{Smette2015}
{Smette} A.,  et~al., 2015, \mn@doi [\aap] {10.1051/0004-6361/201423932}, \href {https://ui.adsabs.harvard.edu/abs/2015A&A...576A..77S} {576, A77}

\bibitem[\protect\citeauthoryear{{Sneppen}}{{Sneppen}}{2023}]{Sneppen2023_bb}
{Sneppen} A.,  2023, \mn@doi [\apj] {10.3847/1538-4357/acf200}, \href {https://ui.adsabs.harvard.edu/abs/2023ApJ...955...44S} {955, 44}

\bibitem[\protect\citeauthoryear{{Sneppen} \& {Watson}}{{Sneppen} \& {Watson}}{2023}]{Sneppen2023_yttrium}
{Sneppen} A.,  {Watson} D.,  2023, \mn@doi [\aap] {10.1051/0004-6361/202346421}, \href {https://ui.adsabs.harvard.edu/abs/2023A&A...675A.194S} {675, A194}

\bibitem[\protect\citeauthoryear{{Sneppen}, {Watson}, {Bauswein}, {Just}, {Kotak}, {Nakar}, {Poznanski}  \& {Sim}}{{Sneppen} et~al.}{2023a}]{Sneppen2023}
{Sneppen} A.,  {Watson} D.,  {Bauswein} A.,  {Just} O.,  {Kotak} R.,  {Nakar} E.,  {Poznanski} D.,   {Sim} S.,  2023a, \mn@doi [\nat] {10.1038/s41586-022-05616-x}, \href {https://ui.adsabs.harvard.edu/abs/2023Natur.614..436S} {614, 436}

\bibitem[\protect\citeauthoryear{{Sneppen}, {Watson}, {Poznanski}, {Just}, {Bauswein}  \& {Wojtak}}{{Sneppen} et~al.}{2023b}]{Sneppen2023_H0}
{Sneppen} A.,  {Watson} D.,  {Poznanski} D.,  {Just} O.,  {Bauswein} A.,   {Wojtak} R.,  2023b, \mn@doi [\aap] {10.1051/0004-6361/202346306}, \href {https://ui.adsabs.harvard.edu/abs/2023A&A...678A..14S} {678, A14}

\bibitem[\protect\citeauthoryear{{Tanvir} et~al.,}{{Tanvir} et~al.}{2017}]{Tanvir2017}
{Tanvir} N.~R.,  et~al., 2017, \mn@doi [\apjl] {10.3847/2041-8213/aa90b6}, \href {https://ui.adsabs.harvard.edu/abs/2017ApJ...848L..27T} {848, L27}

\bibitem[\protect\citeauthoryear{{Tarumi}, {Hotokezaka}, {Domoto}  \& {Tanaka}}{{Tarumi} et~al.}{2023}]{Tarumi2023}
{Tarumi} Y.,  {Hotokezaka} K.,  {Domoto} N.,   {Tanaka} M.,  2023, \mn@doi [arXiv e-prints] {10.48550/arXiv.2302.13061}, \href {https://ui.adsabs.harvard.edu/abs/2023arXiv230213061T} {p. arXiv:2302.13061}

\bibitem[\protect\citeauthoryear{{Vieira}, {Ruan}, {Haggard}, {Ford}, {Drout}, {Fern{\'a}ndez}  \& {Badnell}}{{Vieira} et~al.}{2023}]{Vieira2023}
{Vieira} N.,  {Ruan} J.~J.,  {Haggard} D.,  {Ford} N.,  {Drout} M.~R.,  {Fern{\'a}ndez} R.,   {Badnell} N.~R.,  2023, \mn@doi [\apj] {10.3847/1538-4357/acae72}, \href {https://ui.adsabs.harvard.edu/abs/2023ApJ...944..123V} {944, 123}

\bibitem[\protect\citeauthoryear{{Vieira}, {Ruan}, {Haggard}, {Ford}, {Drout}  \& {Fern{\'a}ndez}}{{Vieira} et~al.}{2024}]{Vieira2023arXiv}
{Vieira} N.,  {Ruan} J.~J.,  {Haggard} D.,  {Ford} N.~M.,  {Drout} M.~R.,   {Fern{\'a}ndez} R.,  2024, \mn@doi [\apj] {10.3847/1538-4357/ad1193}, \href {https://ui.adsabs.harvard.edu/abs/2024ApJ...962...33V} {962, 33}

\bibitem[\protect\citeauthoryear{{Watson} et~al.,}{{Watson} et~al.}{2019}]{Watson2019}
{Watson} D.,  et~al., 2019, \mn@doi [\nat] {10.1038/s41586-019-1676-3}, \href {https://ui.adsabs.harvard.edu/abs/2019Natur.574..497W} {574, 497}

\bibitem[\protect\citeauthoryear{{Waxman}, {Ofek}, {Kushnir}  \& {Gal-Yam}}{{Waxman} et~al.}{2018}]{Waxman2018}
{Waxman} E.,  {Ofek} E.~O.,  {Kushnir} D.,   {Gal-Yam} A.,  2018, \mn@doi [\mnras] {10.1093/mnras/sty2441}, \href {https://ui.adsabs.harvard.edu/abs/2018MNRAS.481.3423W} {481, 3423}

\bibitem[\protect\citeauthoryear{{Wu}, {Barnes}, {Mart{\'\i}nez-Pinedo}  \& {Metzger}}{{Wu} et~al.}{2019}]{Wu2019}
{Wu} M.-R.,  {Barnes} J.,  {Mart{\'\i}nez-Pinedo} G.,   {Metzger} B.~D.,  2019, \mn@doi [\prl] {10.1103/PhysRevLett.122.062701}, \href {https://ui.adsabs.harvard.edu/abs/2019PhRvL.122f2701W} {122, 062701}

\makeatother
\end{thebibliography}

\setcounter{section}{1}
\setcounter{equation}{0}
\setcounter{figure}{0}
\renewcommand{\thesection}{Appendix \arabic{section}}
\renewcommand{\theequation}{A.\arabic{equation}}
\renewcommand{\thefigure}{A.\arabic{figure}}


\end{document}